%% file: main.tex
\documentclass[letterpaper,twocolumn,10pt]{article}
\usepackage{usenix-2020-09}

\usepackage{tikz}
\usepackage{amsmath}

\input{macros}
\input{packages}

\begin{document}

\date{}

\title{\Large \bf Attacking Autonomous Driving Agents with Adversarial Machine Learning:\\A Holistic Evaluation with the CARLA Leaderboard}

\author{
{\rm Henry Wong}\\
Carnegie Mellon University
\and
{\rm Clement Fung}\\
Carnegie Mellon University
\and
{\rm Weiran Lin}\\
Carnegie Mellon University
\and
{\rm Karen Li}\\
Carnegie Mellon University
\and
{\rm Stanley Chen}\\
Carnegie Mellon University
\and
{\rm Lujo Bauer}\\
Carnegie Mellon University
} 

\maketitle

\begin{abstract}
To autonomously control vehicles, driving agents use outputs from a combination of 
machine-learning (ML) models, controller logic, and custom modules.
Although numerous prior works have shown that adversarial examples can mislead ML models used  
in autonomous driving contexts, it remains unclear if these attacks 
are effective at producing harmful driving actions 
for various agents, environments, and scenarios.

To assess the risk of adversarial examples to autonomous driving,
we evaluate attacks against a variety of driving agents,  
rather than against ML models in isolation.  
To support this evaluation, we leverage CARLA, an urban driving simulator, 
to create and evaluate adversarial examples.
We create adversarial patches designed to stop or steer driving agents,
stream them into the CARLA simulator at runtime, and evaluate them 
against agents from the CARLA Leaderboard, a public repository of best-performing 
autonomous driving agents from an annual research competition.
Unlike prior work, we evaluate attacks against autonomous driving systems 
without creating or modifying any driving-agent code and against 
all parts of the agent included with the ML model.

We perform a case-study investigation of 
two attack strategies against three open-source driving agents from the CARLA Leaderboard
across multiple driving scenarios, lighting conditions, and locations.
Interestingly, we show that, although some attacks can successfully 
mislead ML models into predicting erroneous stopping or steering commands,
some driving agents use modules, such as PID control or GPS-based rules, that 
can overrule attacker-manipulated predictions from ML models.

\end{abstract}

\input{docbody}

\bibliographystyle{plain}
\bibliography{biblio}

\appendix
\input{appendix}

\end{document}

%% file: macros.tex
\newcommand{\secref}[1]{Section~\ref{#1}}

\newcommand{\figref}[1]{Figure~\ref{#1}}

\newcommand{\tabref}[1]{Table~\ref{#1}}

\newcommand{\cmark}{\ding{51}}%
\newcommand{\xmark}{\ding{55}}%

\newcounter{finding}
\makeatletter
\newenvironment{finding}[1]
{
    \refstepcounter{finding}
    \protected@edef\@currentlabelname{#1}
    \noindent
    {
    \noindent
        \begin{tcolorbox}
            \textbf{Finding \thefinding}: #1
        \end{tcolorbox}
    }
}
\makeatother

%% file: packages.tex
\usepackage{multirow}
\usepackage{textcomp}
\usepackage{graphicx}
\usepackage{subcaption}
\usepackage{booktabs}
\usepackage{placeins}
\usepackage{makecell}
\usepackage{xurl}

\usepackage{tikz}
\usepackage{amsmath}
\usepackage{amssymb}
\usepackage{pifont}

\usepackage[english]{babel}
\usepackage{xspace}
\usepackage[T1]{fontenc}

\usepackage{csquotes}
\usepackage{tcolorbox}

%% file: docbody.tex
\section{Introduction}
\input{intro}

\section{Background}
\input{background}

\section{Methodology}
\input{methods}

\section{Experimental setup}
\input{setup}

\section{Results of our case study}
\input{results}

\section{Discussion and future work}
\input{futurework}

\section{Conclusion}
\input{conc}

%% file: intro.tex
\label{sec:intro}

Autonomous vehicles are becoming increasingly popular.
These vehicles are controlled by driving agents 
that use machine learning (ML) models 
to predict driving commands based on a vehicle's surrounding environment.
However, ML models are vulnerable to adversarial examples, 
including when physically introduced in driving environments~\cite{sato2021dirty,zhang2022eval,eykholt2018robust}.
Prior work has shown that an attacker can physically alter a vehicle's environment with crafted perturbations 
(e.g., adding markings to the road~\cite{sato2021dirty,boloor2020attack} or street signs~\cite{wang2023does,patel2022billboards}) 
to mislead ML models and cause harmful driving outcomes.
These findings have led to a general concern about the safety of using ML for autonomous driving~\cite{cummings2021rethinking}.

To assess the threat of adversarial examples to autonomous vehicles, 
we must evaluate attacks (i) on systems that are similar to how ML models are used for autonomous vehicles in practice and 
(ii) on scenarios that are systematically defined and reproducible. 
We argue that prior works fall short of meeting these requirements.
First, some prior works only evaluate and measure attack success with ML models in isolation~\cite{wu2020physical,liang2025physical,cheng2024fusion,Nesti22,Nesti24},  
whereas modern autonomous-driving agents use ML as only part of a larger, \emph{end-to-end} pipeline
which combines inputs from multiple modalities (e.g., RGB, GPS, LiDAR) 
and processes ML-model predictions with additional modules for vehicular control.
As a result, even if an attack consistently misleads an ML model into predicting incorrect steering commands, 
the vehicle may not actually perform these steering commands due to the influence of these additional modules. 
Second, other prior works evaluate with fully realized systems and in specific environments~\cite{cao2021invisible,zhu24fusionattack,wang2023does}. 
In addition to the autonomous-driving agent itself, the authors of these works often must also implement the additional software, sensors, and vehicle required for evaluation.
These works also design specific driving scenarios for evaluating their attacks (e.g., building a course on a university campus). 
As a result, the findings from these evaluations are specific to one particular instantiation of a self-driving vehicle and environment that are not generally available to the public.
Because these systems and environments are highly specific and costly to implement, 
it is difficult to reproduce these evaluations and understand how well attacks would translate to other agents.

Thus, a number of important questions remain open for designing more effective attacks and defenses for autonomous driving.
When attacking driving agents with adversarial examples, 
it remains unclear how well previously proposed attack strategies work when modified for contexts different from their original evaluation: 
when used to attack different driving agents, when used for a different attack objective, or when placed in different environment.
When defending driving agents from adversarial examples, it remains unclear which aspects of autonomous-driving agents, 
such as an agent's configuration or the choice of which driving module is used, 
contribute to maintaining safe driving actions in the presence of adversarial examples.

To support holistic, systematic, and reproducible evaluations, we leverage CARLA~\cite{dosovitskiy2017carla}, 
a popular driving simulator, and the CARLA Leaderboard~\cite{carlaleaderboard}, 
a benchmark for evaluating publicly submitted autonomous-driving agents as part of an annual research competition.
We design attacks against open-source agents submitted to the CARLA Leaderboard 
\emph{without making any additions or modifications to the agent or the driving scenarios}--evaluating with 
the submitted ML model weights, submitted agent code and configuration parameters, and common driving scenarios.
We evaluate attack success by crafting adversarial examples, streaming them into the CARLA simulator against a driving agent at runtime,  
and using the metrics for success as defined in the CARLA Leaderboard (e.g., route completion rate, lane infractions). 

To demonstrate the importance of holistically evaluating adversarial attacks on autonomous driving agents, 
we perform attacks with two objectives (stopping and steering) 
against three driving agents (NEAT~\cite{chitta2021neat}, TCP~\cite{wu2022trajectory}, Rails~\cite{chen2021learning}) 
to make comparisons across agents and attack strategies and to analyze the impacts of various attacker and agent configurations.
We are able to show that strategies that use adversarial examples to erroneously stop a vehicle often generalize across agents, 
whereas other attack objectives require more agent-specific customization.   
For instance, to cause a vehicle to steer erroneously, attacks must first mislead the underlying ML model and
then also bypass mitigations from agent-specific PID controllers and GPS-based rules.

Drawing inspiration from the collaborative and competitive nature of the CARLA Leaderboard, 
we propose developing a new leaderboard for evaluating adversarial examples on agents in CARLA, 
which will serve as a red-and-blue-team platform for developing stronger attacks and 
more robust agents in standardized environments.

%% file: background.tex
\label{sec:background}

In this section, we describe the background for our work. 
We describe autonomous driving agents in \secref{back:e2e},
CARLA in \secref{back:carla},
and discuss prior work in adversarial ML for autonomous driving in \secref{back:attacks}. 

\subsection{Autonomous driving agents}
\label{back:e2e}

As opposed to an ML model that performs a task associated with autonomous driving 
(e.g., lane detection~\cite{wang2018lanenet}, traffic sign recognition~\cite{flores2024traffic}, 
vehicle recognition~\cite{bai2022transfusion}, route planning~\cite{althoff2017commonroad}), 
an autonomous driving agent predicts driving actions (e.g., steer left, apply brake) 
based on sensor inputs from a vehicle's environment~\cite{chen2024end}.
Autonomous driving agents often use multiple ML models and often use non-ML-based modules, such as PID control and GPS-based rules, to output control commands. 
In this work, we evaluate attacks against autonomous driving agents, 
rather than evaluating against individual ML models used for 
autonomous-driving tasks.

\subsection{CARLA simulator and leaderboard}
\label{back:carla}

In this work, we use the CARLA 0.9.10 simulator~\cite{dosovitskiy2017carla,carlaorg}, an open-source, urban driving simulator.
Although other high-fidelity simulators and software stacks for autonomous driving have been used in 
prior work~\cite{rong2020lgsvl,shen2022pass,kato2018autoware,fan2018baidu}, open-source models for these simulators are rare and are not widely shared.
To evaluate attacks with these simulators, we would need to design our own custom driving agents, which is costly and requires choosing ML models and techniques that are the state-of-the-art, 
which are currently unclear.

In contrast, we use the CARLA Leaderboard 1.0~\cite{carlaleaderboard}, a public repository of end-to-end driving agents that have been submitted to an annual competition since 2019.
To evaluate autonomous driving agents, the CARLA Leaderboard 1.0 API provides a standard interface that agents must use to read sensor information from their driving environment 
(e.g., images, LiDAR, GPS) and send driving commands to a vehicle (e.g., steering and acceleration values).
Many agents that have been submitted to the CARLA Leaderboard (shown in \tabref{table:agents} and described in more detail in \secref{framework:agents})
provide ML model weights and open-source code implementations.
In this work, we evaluate attacks against open-source agents on the CARLA Leaderboard to compare attack outcomes across agents and attack strategies.

\subsection{Adversarial ML for autonomous driving}
\label{back:attacks}

Prior work has developed several techniques based on adversarial ML to attack ML models.
Attackers with (black-box or white-box) access to a target ML model can optimize and generate perturbations for a defined attack objective.
These perturbations are then applied to the target model's inputs to mislead the model into making predictions that meet the defined attack objective~\cite{szegedy2013intriguing}.
Common attack objectives include misclassifying images~\cite{sharif2016accessorize,eykholt2018robust},
inducing false object detection~\cite{wang2021daedalus}, 
and evading detection algorithms~\cite{biggio2013evasion,lucas2021malware}.

In the context of self-driving vehicles, prior work uses adversarial examples to mislead 
lane detection models~\cite{sato2021dirty,jing2021tricking}, evade object detection~\cite{cheng2024fusion,Nesti22,Nesti24}, or mislead traffic sign classifiers~\cite{wang2023does,eykholt2018robust}; 
however, these works only attack individual models and therefore do not measure how driving-agent modules can affect attack outcomes.
Some prior works attack ML models that predict autonomous-driving actions~\cite{boloor2020attack,yang2020attack,patel2022billboards},
but these attacks are evaluated with a model developed for CARLA 0.8 (i.e., an older version)~\cite{codevilla2018cil}. 
We observe that modern driving agents developed for CARLA 0.9 and the CARLA Leaderboard 1.0 API 
combine ML models with other modules, such as PID control and GPS-based rules. 
In this work, we evaluate adversarial examples designed to mislead autonomous driving agents 
into performing erroneous driving actions.

%% file: methods.tex
\label{sec:framework}

\input{tables/agents-table.tex}

\begin{figure*}[t!]
    \centering
    \includegraphics[width=0.95\linewidth]{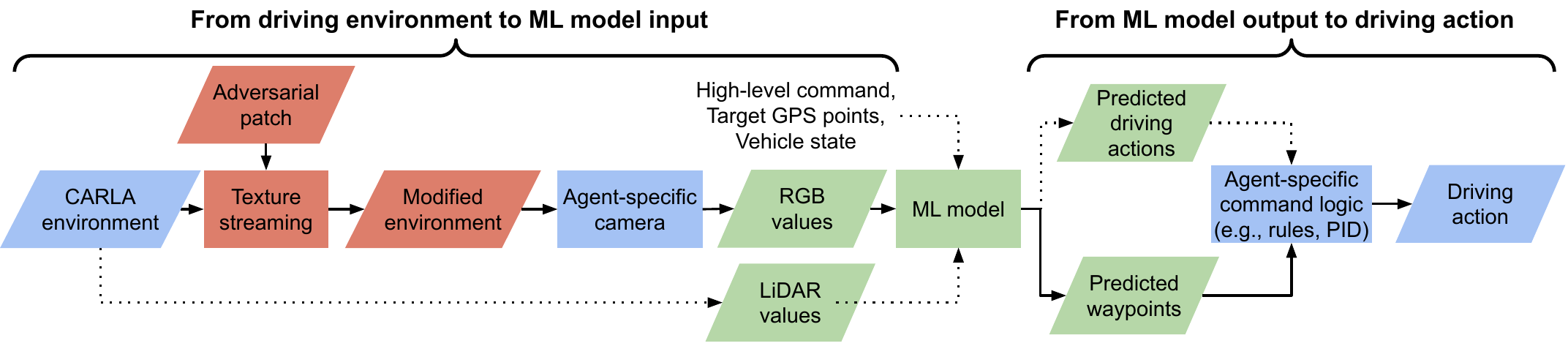}
    \caption{We show the various parts of autonomous driving agents based on the CARLA Leaderboard API. 
        These parts include ML models (in green) and other modules used prior to ML model inputs or after ML model outputs (shown in blue). 
        We evaluate agents that use RGB images for model input and predict waypoints as model output (shown in solid lines); 
            however, we show other commonly used input and output modalities in this diagram (shown in dotted lines). 
            In red, we show our process for inserting adversarial examples into CARLA.}
    \label{fig:framework}
\end{figure*}

In this section, we describe our methodology for evaluating adversarial examples against driving agents with CARLA.
We first analyze and describe important characteristics of agents submitted to the CARLA Leaderboard (\secref{framework:agents}). 
Next, we discuss the ethics considerations of our work (\secref{framework:ethics}).
Finally, we describe our threat model (\secref{framework:tm})
and methodology for creating and evaluating adversarial examples in CARLA (\secref{framework:attacks}). 

\subsection{Characterizing agents submitted to the CARLA Leaderboard}
\label{framework:agents}

We inspect the descriptions and code implementations of all open-source agents that have been submitted 
to the CARLA Leaderboard
since 2019\footnote{In 2025, the CARLA Leaderboard migrated to EvalAI (\url{https://eval.ai/web/challenges/challenge-page/2098/overview}), 
and historical records of agent performance were removed. Our work is based on the best-performing agents that were submitted from 2019--2023, prior to the migration to EvalAI.}
and characterize their input and output modalities in \tabref{table:agents}.

Early submissions to the CARLA leaderboard often use RGB inputs to directly predict control commands~\cite{codevilla2019cilrs,toromanoff2020marln,chen2021learning}. 
As teams continue to build and innovate on prior submissions, 
we observe that agents have begun to coalesce more advanced strategies, 
such as using ML models to predict waypoints rather than control commands 
(e.g., NEAT~\cite{chitta2021neat} and TCP~\cite{wu2022trajectory}) and 
using ML models that perform sensor fusion across data from RGB, GPS, and LiDAR (e.g., LAV~\cite{chen2022learning} and Transfuser~\cite{chitta2023transfuser}).
Based on our observations, we define a general structure for driving agents submitted to the CARLA Leaderboard, shown in \figref{fig:framework}.
We show parts of the pipeline that are direct inputs and outputs of the target ML model (shown in green), 
as well as auxiliary parts of the pipeline (shown in blue) used to predict driving actions.

\paragraph{Input modalities.}
We first describe the various input modalities used by driving agents (left side of \figref{fig:framework}).
The CARLA Leaderboard API provides several inputs including RGB camera, LiDAR data, GPS, and high-level commands. 
We characterize agents by their input modalities in \tabref{table:agents}.
All agents use RGB images from the vehicle's perspective.
Some agents also use LiDAR to construct a point-cloud image of objects surrounding the vehicle.
Some agents also use GPS; target points define routes in CARLA scenarios, and these agents compare the vehicle's GPS location with GPS targets to adjust driving actions. 
Finally, some agents use high-level commands provided by the CARLA Leaderboard scenario---a categorical input with values such as ``Lane Follow'' or ``Turn Right''.

\paragraph{Output modalities.}
We next describe the output modalities used by ML models in driving agents. 
Many agents use ML models to predict driving actions (i.e., steer, brake, and throttle) at each time step, which are directly sent as commands to the vehicle.
Alternatively, many driving agents instead predict a forecasted list of waypoints.
These waypoints define a desired future path for the vehicle, representing both speed and direction.
To convert these waypoints into control commands, driving agents use PID (proportional, integration, derivative) controllers, which serve as low-level planning modules~\cite{chen2020lbc,chitta2021neat}. 
At each time step, PID controllers use waypoints as input to produce a steering and throttle command, which are then sent the vehicle.

\paragraph{Agent-specific rules.}
To mitigate errors, some agents use rules that can explicitly override the ML model predictions.
For instance, if the predicted waypoints point too far from the next GPS target point, the NEAT and TCP agents will instead use the direction of the next GPS target as the steering angle~\cite{chitta2021neat,wu2022trajectory}.
These rules are configured for deviations and distances specific to each agent.
Prior work has shown that GPS-based rules can cause driving-agent errors in benign driving scenarios~\cite{jaeger2023biases},
and in this work we explore how these rules can affect agent commands in attack scenarios.

\subsection{Ethics considerations}
\label{framework:ethics}

To the best of our knowledge, none of the agents attacked in our work are used in any real, deployed systems, so we believe the harm of our work is minimal. 
As a benefit, our work investigates attack strategies against publicly submitted agents to better understand how design factors affect robustness to adversarial examples,
with the end goal of designing driving agents that are more robust and safe. 

\subsection{Threat model}
\label{framework:tm}

In this section, we describe our attacker threat model.

\paragraph{Goals.}
We assume that the attacker's goal is to manipulate the movement of the target vehicle
by causing the vehicle to stop or causing the vehicle to exit its driving lane.
We describe formal definitions of these attack goals in \secref{framework:attacks}.  

\paragraph{Knowledge.} We assume that the attacker has access to the parameters of 
a ML model (i.e., white-box access) which processes RGB in the target vehicle's 
driving agent. For instance, the attacker may know that an open-source ML model is used in 
their target vehicle's driving agent and downloads the weights of this ML model 
from a public repository. 
However, we assume that the attacker \emph{does not} have access to the other parts 
of the agent, such as other sensor inputs, PID control, or rules, and cannot modify 
any part of the driving agent's ML model, code, or configuration.

\paragraph{Capabilities.}
We assume the target driving agent will autonomously navigate to a public location controlled by the attacker. 
To execute the attack, we assume that the attacker can print and deploy a patch on a surface in this public location, such as on the road, a sign, or a building. 
Once placed, we assume that the adversary is unable to modify the patch further (i.e., the attack is non-adaptive). 
We also assume that the attacker is unable to modify other data modalities; in particular, they cannot modify the GPS or LiDAR readings sent to the target driving agent.
Since we assume that driving agents operate autonomously without human involvement,
the attacker does not need to consider strategies to avoid human perception 
(e.g., making their patch imperceptible).

\subsection{Attack methodology}
\label{framework:attacks}

We describe our methodology for generating and evaluating adversarial patches for our attack.

\paragraph{Defining the attack goals.}
We propose two separate attack goals---stopping and steering. 
Given an ML model that predicts steering ($s_t \in [-1, 1]$), throttle ($a_t \in [0,\infty])$, and braking ($b_t \in [0, 1]$) at time $t$, we define the attack loss $A(t)$ as the total braking for a stopping attack ($A(t) = 1 - b_t$)
and as the weighted sum of squared differences from a target speed $v^*$ and steering angle $s^*$ for a steering attack ($A(t) = \delta || v_t - v^*||^2 + \epsilon || s_t - s^*||^2$).

Some ML models instead predict a list of waypoints $(x_0, y_0), ..., (x_n, y_n)$, 
where the y-axis is the forward direction and the vehicle's position is at $(0,0)$. 
For these models, we define the attack loss for a stopping attack as the L2 norm of the waypoints ($A(t) = \sum_{i=1}^n \| (x_i, y_i) \|^2$). 
We define the attack loss for a steering attack as the weighted sum of speed loss 
(distance between the entire trajectory magnitude and a target speed $v^*$) and steering loss (distance between the waypoint aim angle and the target steering angle $s^*$): 
\begin{align*} 
    V(t) &= \| (\sum_{i=0}^{n - 1} \|(x_{i+1}, y_{i+1}) - (x_i, y_i)\|^2) - v^* \|^2 \\
    S(t) &= \frac{1}{n}\sum_{i=0}^n\| \left(\frac{2}{\pi}\left(\frac{\pi}{2} - \tan^{-1}\left(\frac{-y_i}{x_i}\right)\right) - s^*\right)\|^2 \\
    A(t) &= \delta V(t) + \epsilon S(t)
\end{align*}

\paragraph{Generating adversarial patches.}
To use our attack loss to generate adversarial patches, 
we first collect images of driving scenes in the attacker-chosen location.
Similar to prior work~\cite{patel2022billboards}, we drive the target agent 
through the target location several times in the CARLA simulator to collect its input RGB and vehicle state.

Once these images are collected, we then insert our candidate patches into these images.
To faithfully represent a patch's appearance to the vehicle's camera in CARLA, 
(i) we use a projection algorithm to map a patch to its size and orientation within the image, 
(ii) we perturb the patch's RGB values to account for simulator lighting effects, 
and (iii) we blur patches to account for lowered camera resolutions.
We provide additional details in \secref{sec:projection}.

Given a set of images of driving scenes and patch mappings, we search for patch values 
that minimize the attack loss over the weights of the target ML model.
We optimize the patch $p$, over our set of images $X$, for model $\mathrm{M()}$.
To help generate patches that can be rendered in the CARLA simulator, 
we use methods from prior work~\cite{sharif2016accessorize} 
and include regularization terms for print-ability, total variation, and saliency.
Our overall minimization objective is:
\[ \min_p \sum_{x \in X} \alpha \mathrm{A}( \mathrm{M}(x + p) ) + \beta \mathrm{NPS}(p) + \gamma \mathrm{TV}(p) + \eta \mathrm{SAL}(p) \]

\paragraph{Evaluating attack success.}
To evaluate adversarial patches against driving agents in CARLA, 
we stream the patch into the CARLA environment at test time 
using the texture streaming API (shown in red in \figref{fig:framework}).
Next, we execute the driving scenario in the CARLA Leaderboard, which uses commands from the driving agent to control a vehicle through the manipulated scene.
We evaluate with the built-in driving metrics from the CARLA Leaderboard, which record 
lane infractions, route deviation, and route completion~\cite{carlaleaderboard}. 
For further analysis, we also instrument the CARLA Leaderboard to record auxiliary values from CARLA, 
such as the vehicle's speed, steering, throttle, and GPS position, as well as inputs and outputs of the target ML model.

\subsection{Patch projection methodology}
\label{sec:projection}

In this section, we describe our methodology for projecting patches into RGB images;
we perform these projections to help ensure that we optimize patches using images 
that correctly represent how an adversarial patch would appear from the perspective 
of an approaching vehicle in the CARLA simulator.

\paragraph{Adjusting for size and orientation.}
Given the specified global orientation, which includes both position and rotation, along with patch-specific parameters, 
such as physical size and resolution, we transform a 2-dimensional rectangular source image of the patch into 
its corresponding appearance from the perspective of all cameras of the vehicle. 
Since this projection is used during optimization, we utilize PyTorch's perspective transformations, 
allowing gradients to flow through the projection function and enable the optimization of patch values.
Using the patch center and its dimensions, we calculate the locations of the patch corners in 3-dimensional space, relative to the camera. 
Using the camera's intrinsic and extrinsic parameters, we then subsequently project the patch and its associated mask 
onto the image plane using perspective transformations, while also filtering out points that are located behind the camera. 
Overall, projections are crucial for accurately simulating the appearance of each patch 
from various camera angles within the CARLA environment, particularly as optimization occurs at different angles.

\paragraph{Perturbing colors and edges.}
To enhance the consistency of our attack against slight variations in color and resolution, we introduce additional perturbations to the patch values while projecting them onto an image. 
Specifically, we apply varying degrees of color jitter to the source images and Gaussian noise to the input images.

CARLA's realistic lighting effects make optimization challenging, as these effects bleach the colors of patches in daytime settings. 
Although we explored creating a lighting model to estimate the mapping of source colors to their appearance in CARLA, 
we ultimately decided to disable lighting effects on the deployed patch assets. 
This allowed us to concentrate on optimizing the patches for attack success. 
In CARLA, we modified the shading model of the base material to 'Unlit' and set the emissive color to the patch source image (scaled by a factor of 255). 
Although these changes remove realistic lighting effects, minor color discrepancies between the source and real-world patches still exist. 
To enhance robustness and better replicate the patch colors as viewed in the simulator, 
we varied the brightness, contrast, and saturation values per pixel by linearly scaling them with a small, uniformly sampled multiplicative factor.

To simulate the down-sampled resolution quality from cameras and the anti-aliasing effects on distant patches, 
we added Gaussian noise to blur the patches and blend them with the surrounding background from the input images. 
This approach softens the harsh edges between pixels and produces results that more closely resemble the true camera outputs.

%% file: tables/agents-table.tex
\begin{table*}[tb!]
	\centering
	\caption{
        We list all agents that have been submitted to the CARLA Leaderboard that include open-source code.
        For each agent, we list its name, its year of publication, its inputs, and the output modalities used in its primary ML model. We observe that ML models increasingly use LiDAR and predict waypoints, rather than directly predicting control commands from images only.
    }
	\label{table:agents}
	\begin{tabular}{rr cccc cc}
	\toprule
        & & \multicolumn{4}{c}{\textbf{Inputs}} & \multicolumn{2}{c}{\textbf{Outputs}} \\
        \cmidrule(lr){3-6} \cmidrule(lr){7-8}
        \textbf{Name} & \textbf{Year} & RGB & LiDAR & High-level Command & GPS Targets & Waypoints & Control Commands \\
        \midrule

        Transfuser~\cite{chitta2023transfuser}  & 2023 & \cmark & \cmark & \xmark & \cmark       & \cmark & \xmark \\
        InterFuser~\cite{shao2023interfuser}    & 2023 & \cmark & \cmark & \xmark & \cmark       & \cmark & \xmark \\
        LAV~\cite{chen2022learning}             & 2022 & \cmark & \cmark & \cmark & \cmark       & \cmark & \xmark \\

        TCP~\cite{wu2022trajectory}             & 2022 & \cmark & \xmark & \cmark & \cmark       & \cmark & \cmark \\ 
        NEAT~\cite{chitta2021neat}              & 2021 & \cmark & \xmark & \xmark & \cmark       & \cmark & \xmark \\
        Rails~\cite{chen2021learning}           & 2021 & \cmark & \xmark & \cmark & \xmark       & \xmark & \cmark \\

        LBC~\cite{chen2020lbc}                  & 2020 & \cmark & \xmark & \xmark & \xmark       & \cmark & \xmark \\
        IA~\cite{toromanoff2020marln}           & 2020 & \cmark & \xmark & \cmark & \cmark       & \xmark & \cmark \\
        CILRS~\cite{codevilla2019cilrs}         & 2019 & \cmark & \xmark & \cmark & \cmark       & \xmark & \cmark \\

    \bottomrule
	\end{tabular}
\end{table*}

%% file: setup.tex
\label{sec:setup}

\begin{figure}[t!]
    \centering
    \includegraphics[width=0.99\linewidth]{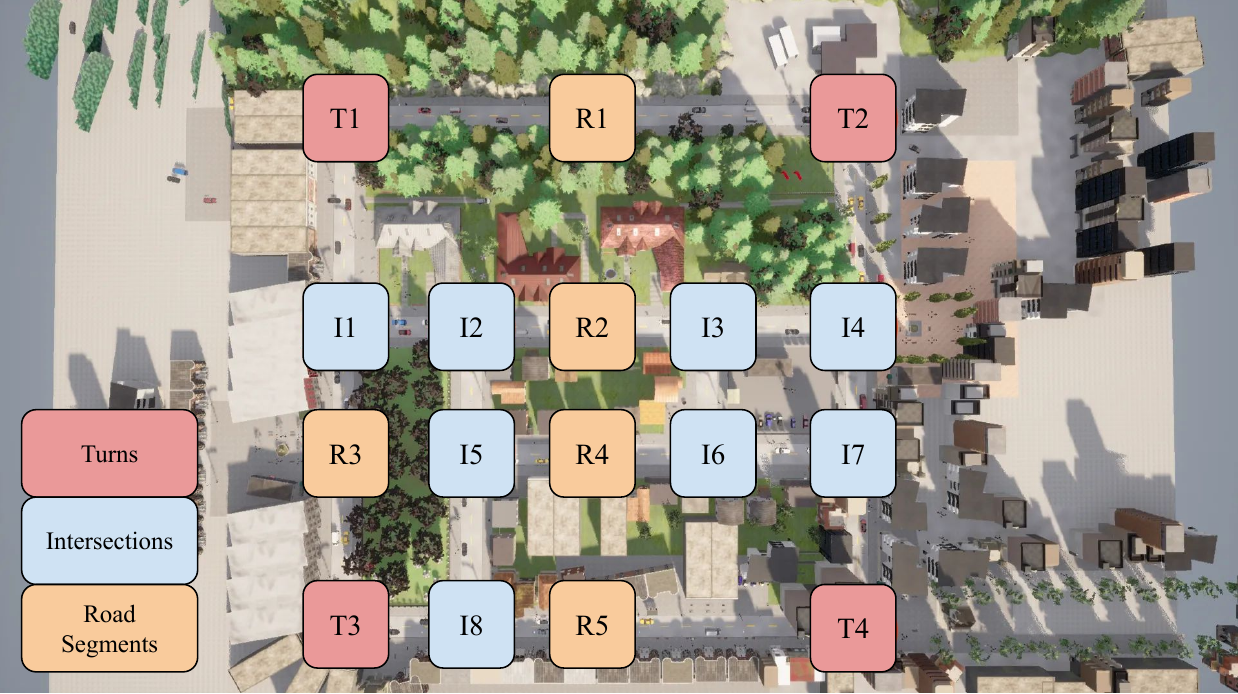}
    \caption{We annotate a map of Town~02 in CARLA, listing all locations where a curved road segment, a straight road segment, or an intersection are found. We number each of these instances with a unique identifier.}
    \label{fig:carla-map}
\end{figure}

For our representative case study, we evaluate attacks on three agents from the CARLA Leaderboard with different output modalities: 
TCP~\cite{wu2022trajectory}, NEAT~\cite{chitta2021neat}, and Rails~\cite{chen2021learning}. 
As shown in \tabref{table:agents}, these three agents use different combinations of output modality and use RGB inputs without sensor fusion.
All of our experiments are performed in Town~2, an urban environment in CARLA.
After identifying all straight, curved, and intersecting road segments in Town~2, we choose two locations for evaluation 
(Road~\#5 and Turn~\#2, shown in \figref{fig:carla-map}). 
These two locations are used in two separate CARLA Leaderboard scenarios;
in benign settings, all three agents successfully guide a vehicle through these scenarios without any infractions.

As described in \secref{framework:attacks}, we drive each agent through each target location several times, record its inputs,
and generate patches for each attack goal using our loss functions. 
We provide additional experimental details in Appendix~\ref{app:setup}.

%% file: results.tex
\label{sec:results}

In this section, we describe our results from performing stopping attacks and steering attacks 
in \secref{results:stopping} and \secref{results:steering} respectively. 
We analyze success and failure cases for three driving agents, based on agent-specific factors.

\begin{figure}[t!]
    \begin{subfigure}{\linewidth}
        \centering
        \includegraphics[width=0.95\linewidth]{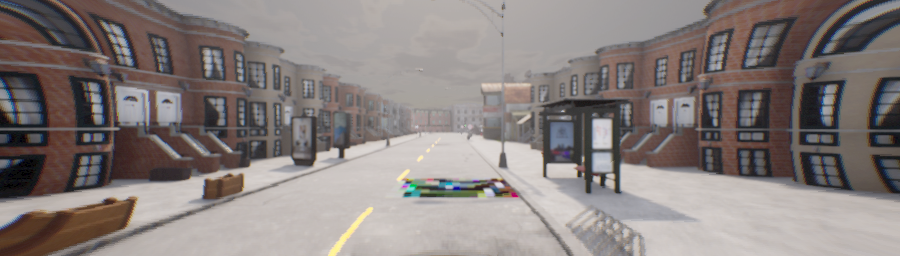}
        \caption{RGB input (frame 70) with the light-optimized patch.}
        \label{fig:stopping-attack-frame}
    \end{subfigure}
    \begin{subfigure}{\linewidth}
        \centering
        \includegraphics[width=0.95\linewidth]{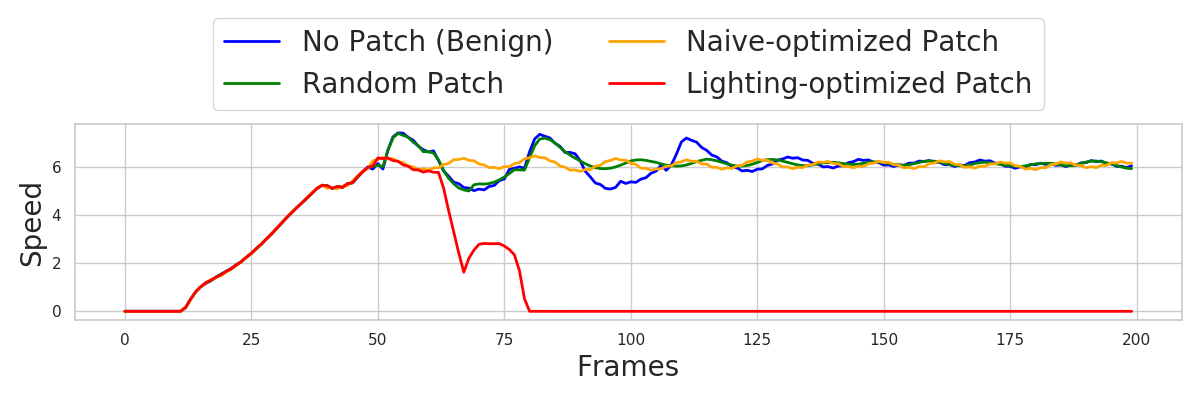}
        \caption{Vehicle speed for each frame in four different evaluation settings. 
            The vehicle only stops when the agent approaches the light-optimized patch (shown in red).}
        \label{fig:stopping-attack-speed}
    \end{subfigure}
    \caption{We generate an adversarial patch that successfully misleads the TCP agent to stop the vehicle, resulting in a route completion failure on the CARLA Leaderboard.}
    \label{fig:stopping-attack}
\end{figure}
\begin{figure}[t!]
    \begin{subfigure}{\linewidth}
        \centering
        \includegraphics[width=0.99\linewidth]{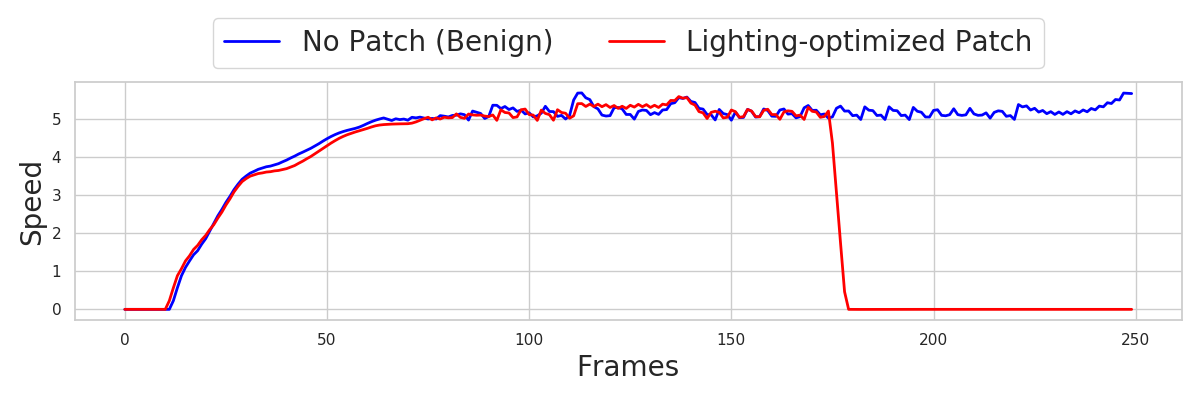}
        \caption{Results of our stopping attack against the Rails agent.}
        \label{fig:stopping-attack-rails}
    \end{subfigure}
    \begin{subfigure}{\linewidth}
        \centering
        \includegraphics[width=0.99\linewidth]{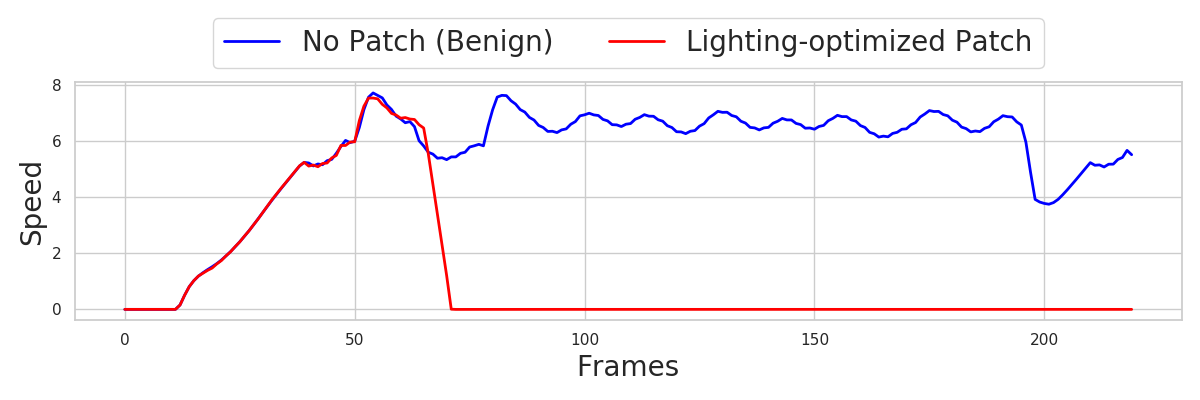}
        \caption{Results of our stopping attack against the NEAT agent.}
        \label{fig:stopping-attack-neat}
    \end{subfigure}
    \caption{We show the speed of the target vehicle when performing our stopping attack against the Rails agent (top) and the NEAT agent (bottom).
        In both cases, the adversarial patch successfully causes the vehicle to stop, resulting in a route completion failure.}
    \label{fig:stopping-attack-others}
\end{figure}

\subsection{Stopping attack}
\label{results:stopping}

We evaluate adversarial examples designed for our stopping attacks 
by placing the adversarial patch on Road \#5 and 
using the target agent to control the vehicle through this location, 
as shown in \figref{fig:stopping-attack-frame}.
\figref{fig:stopping-attack-speed} shows the vehicle's driving speed 
when controlled by TCP, across four different evaluation scenarios. 
We show the driving speed with no patch (i.e., benign scenario) 
and when our patch is placed on the road.

To compare the effects of different patch projection methods, we also show the results 
with a patch of random Gaussian noise, 
and with a patch that is generated without our lighting optimization approach (described in \secref{sec:projection}), 
which we call ``naive-optimized''.
Across all evaluation scenarios, only the lighting-optimized patch 
causes the vehicle to stop, triggering a route completion failure on the CARLA Leaderboard. 
We repeat our evaluation three times and find that each trial yields the same result. 
When lighting optimization is not used during patch optimization, the attack fails. 
However, correcting for errors in color and lighting are not always sufficient; 
in \secref{results:steering} we show that agent-specific rules and controllers 
can mitigate steering attacks.

We find that our attacks are also successful against NEAT and Rails, as shown in \figref{fig:stopping-attack-others}.
In all cases, our lighting-optimized adversarial patches can successfully stop the autonomous vehicle, 
triggering a route completion failure in the CARLA Leaderboard. 

\begin{finding}{By adjusting for lighting, color, and resolution in CARLA, an attacker can generate adversarial patches that are effective at misleading driving agents into stopping the target vehicle.}
\end{finding}

\begin{figure}[t!]
    \begin{subfigure}{\linewidth}
        \centering
        \includegraphics[width=0.95\linewidth]{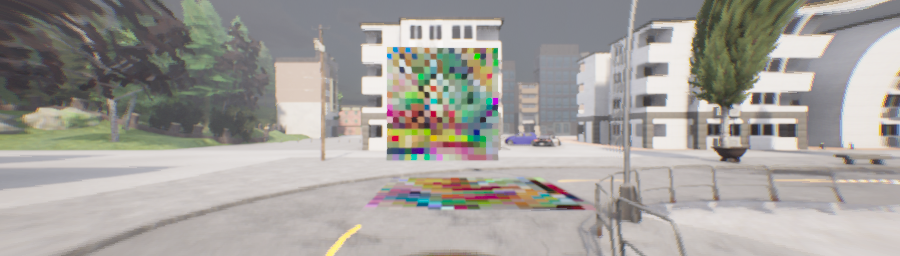}
        \caption{RGB input (frame 120) of the steering attack.}
        \label{fig:steering-attack-frame}
    \end{subfigure}
    \begin{subfigure}{\linewidth}
        \centering
        \includegraphics[width=0.95\linewidth]{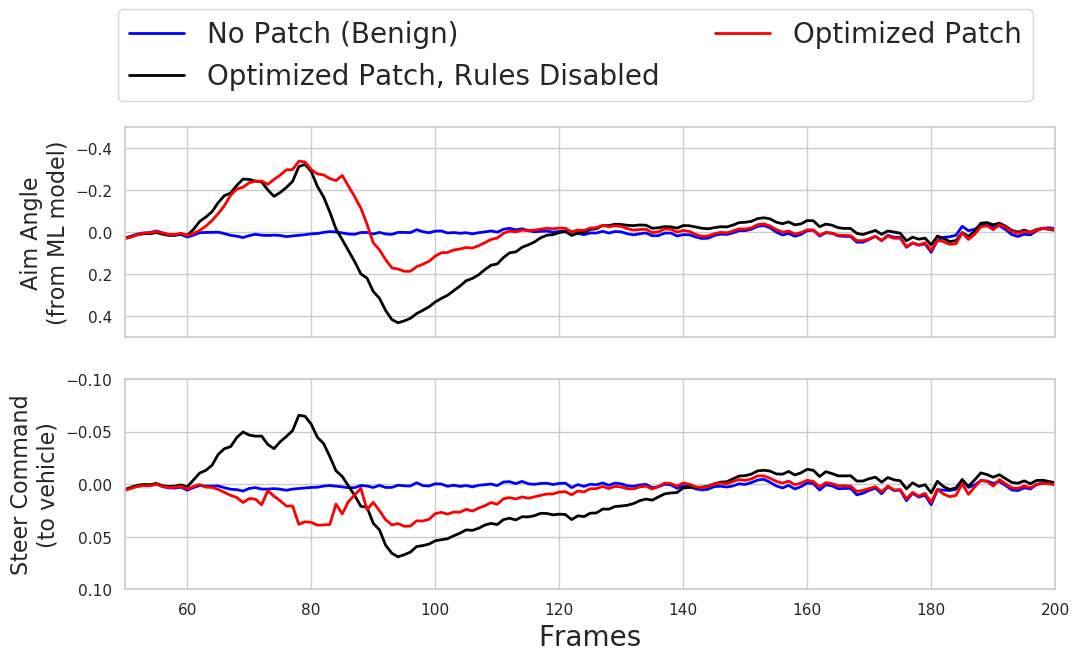}
        \caption{Steering attack on a Road \#5, a straight road segment.}
        \label{fig:steering-attack-aim-1}
    \end{subfigure}
    \begin{subfigure}{\linewidth}
        \centering
        \includegraphics[width=0.95\linewidth]{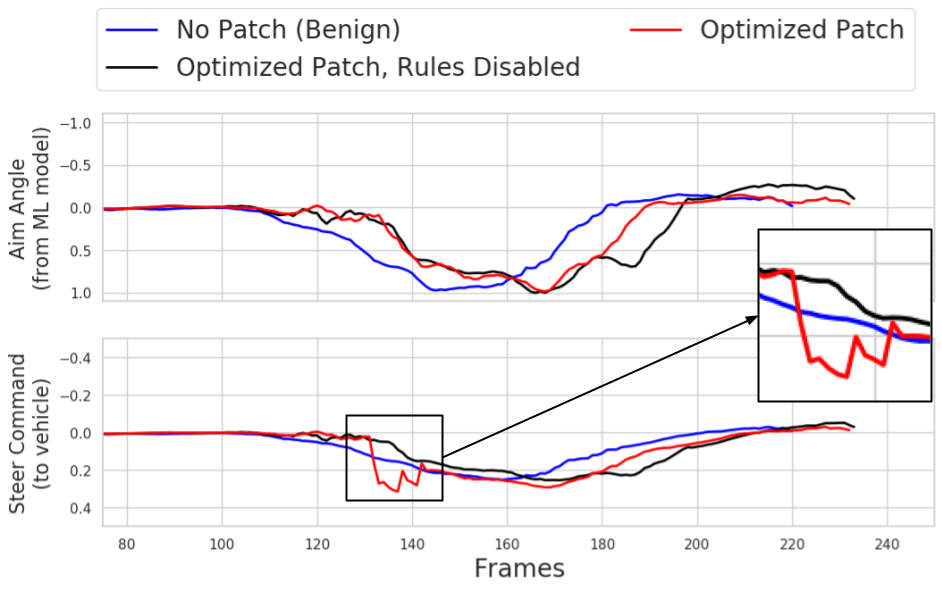}
        \caption{Steering attack on Turn \#2, a curved road segment. 
            We highlight the behavior from frames 120--150: 
                our adversarial patches (in red) induce an aim angle with a weaker right turn, 
                but the corresponding steering command is adjusted by GPS-based rules.}
        \label{fig:steering-attack-aim-2}
    \end{subfigure}
    \caption{We perform steering attacks against TCP on Road \#5 (b) and Turn \#2 (c),
        and we compare the ML model's predicted aim angle (top) to the vehicle's steering command (bottom). 
        In both locations, the adversarial patches manipulate the aim angle (red, top) but not the steering commands (red, bottom). 
        We manually disable TCP's agent-specific rules (black), and find that only then do our attacks succeed.}
    \label{fig:tcp-steering-aim}
\end{figure}

\begin{figure}[t!]
    \centering
    \includegraphics[width=0.99\linewidth]{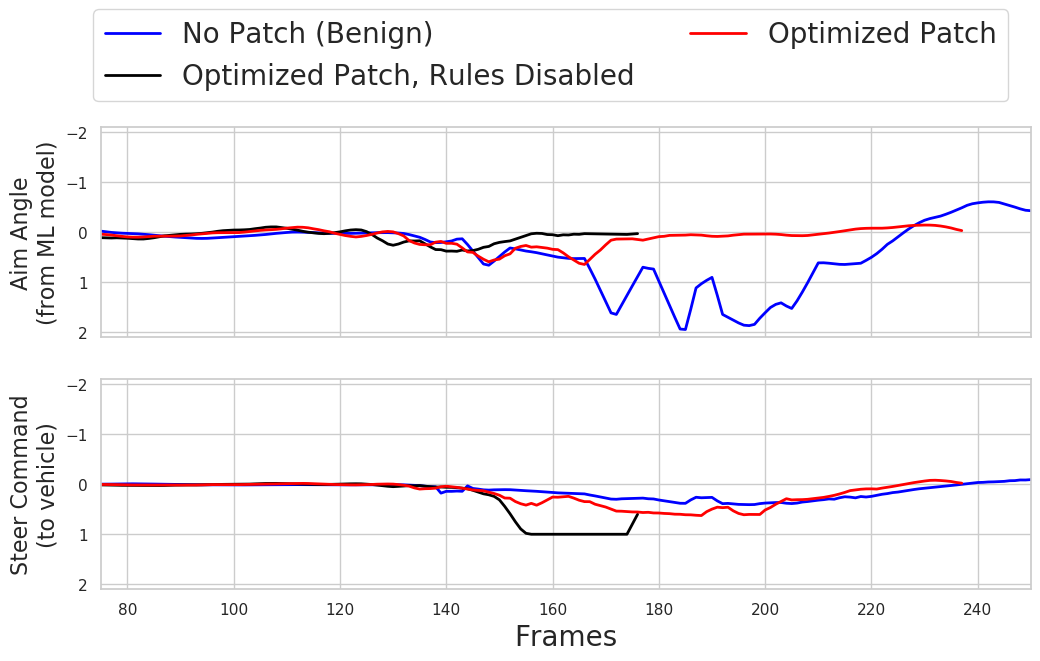}
    \caption{We perform steering attacks against NEAT on Turn \#2. 
        We compare the ML model's predicted aim angle (top) to the vehicle's corresponding steering command (bottom). 
        Although our adversarial patch manipulates the aim angle (red, top), the steering commands follow the benign case (red and blue, bottom). 
        We manually disable agent-specific rules (black), but find that the attack still does not succeed.}
    \label{fig:neat-steering-aim}
\end{figure}

\begin{figure}[t!]
    \begin{subfigure}{\linewidth}
        \centering
        \includegraphics[width=0.99\linewidth]{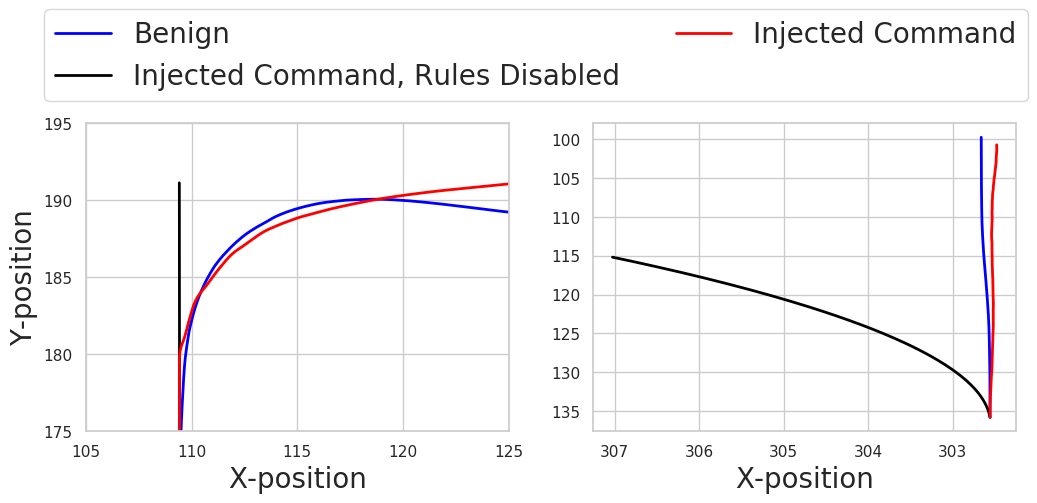}
        \caption{The vehicle GPS position when attacking the NEAT agent.}
        \label{fig:neat-steering-position}
    \end{subfigure}
    \begin{subfigure}{\linewidth}
        \centering
        \includegraphics[width=0.95\linewidth]{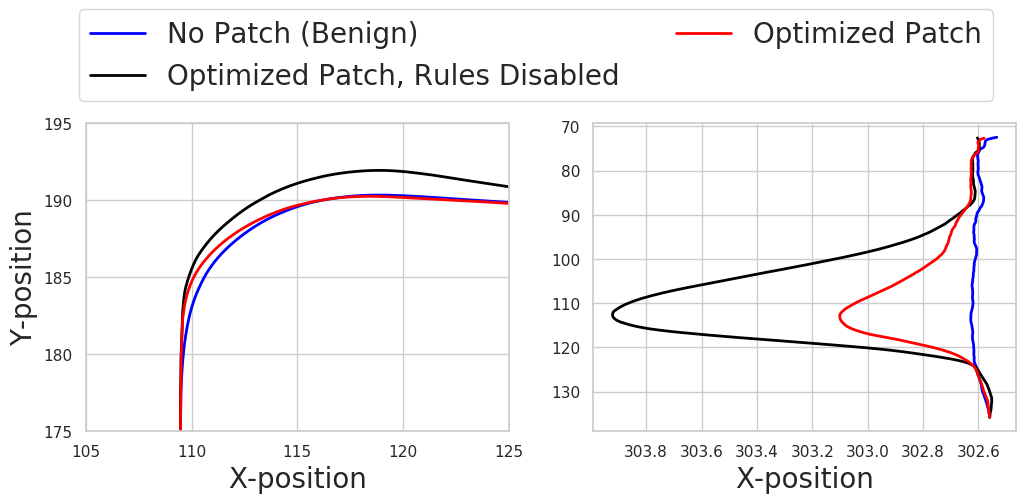}
        \caption{The target vehicle GPS position when attacking the TCP agent.}
        \label{fig:tcp-steering-position}
    \end{subfigure}
    \caption{We show the GPS position of the target vehicle on Turn~\#2 (left) and Road~\#5 (right) when injecting commands to the NEAT (top) and TCP (bottom) agents. 
            Even when the aim angle is manually injected (red), agent-specific rules provide the steering commands.
            Lane infractions can only occur when agent-specific rules are disabled (black).}
        \label{fig:steering-position}
\end{figure}

\subsection{Steering attack}
\label{results:steering}

Next, we describe the results of our steering attacks. 
We evaluate steering attacks on both Road~\#5 and Turn~\#2.
For Road~\#5, we design patches to steer vehicles left into the opposing lane, 
and for Turn~\#2 (i.e., a right turn), 
we design patches to mislead the vehicle into accelerating forward 
and missing the turn.

We initially found that all patches fail to mislead agents.
After further analysis, we found that our adversarial patches can successfully mislead predictions of the ML model, 
but these predictions do not persist through the agent's PID controller and rules. 
\figref{fig:tcp-steering-aim} shows the aim angle (i.e., the reference angle computed from the ML model's predicted waypoints) 
and the steering command sent to the target vehicle for Road~\#5 (top) and Turn~\#2 (bottom) respectively for the TCP agent.
For instance, our adversarial patches induce a left-turning aim angle on Road~\#5 
but the agent does not send left-turning steering commands to the vehicle.
When performing our steering attacks against NEAT, we faced challenges from agent-specific rules and PID controllers, similar to with TCP. 
\figref{fig:neat-steering-aim} compares the aim angle (predicted by the ML model) and the steering command for Turn~\#2. 
We find that, although our adversarial patches cause the model to predict a weaker right turn (in red, on top), 
the steering command sent to the vehicle still follows its benign counterpart. 

For NEAT and TCP, agent-specific rules prevent the attack from inducing erroneous driving actions; to verify this, 
we perform an additional experiment in which we manually set the predicted aim angle to a constant left turn 
(i.e., completely bypassing the prediction from the ML model)
and the vehicle still does not exit the driving lane.

In our final demonstration, we manually disable TCP's agent-specific rules and find that
only then are our adversarial patches successful at triggering a lane infraction from the CARLA Leaderboard 
(shown in black in \figref{fig:tcp-steering-aim}). 
We perform the same experiment with NEAT, shown in \figref{fig:neat-steering-aim}; 
we disable NEAT's agent-specific rules (in black), but find that the agent still sends right-turning steering commands to the vehicle, likely due to the nature of its PID controller. 
To measure the impact of PID control and agent-specific rules, we manually set to predicted aim angle to a constant value. 
We find that agent-specific rules prevent the injected aim angle from controlling the vehicle, and the PID controller permits the injected command. 
\figref{fig:steering-position} shows the corresponding GPS positions of target vehicle when performing our steering attacks. 
We show that the position and trajectory of the vehicle is sufficiently modified, triggering a lane infraction violation in the CARLA Leaderboard.

As a final remark, we analyze the navigation routes used across all CARLA Leaderboard driving scenarios and find that, due to GPS-based rules alone,
steering attacks are impossible on TCP for at least 30\% of locations in Town~2. 

\begin{finding}{Using adversarial patches to manipulate the steering of driving agents is difficult or even impossible in some situations.
    Some agents use rules or PID controllers that can overrule the predictions of ML models.}
\end{finding}

%% file: futurework.tex
\label{sec:futurework}

In this section, we discuss proposals for future work 
based on our findings.

\paragraph{Exploring cross-modal attacks on sensor-fusion models.}
In our case study, we only explored attack strategies that manipulate the RGB image domain.
However, we observe that agents are increasingly using models for autonomous driving 
that perform sensor fusion across RGB and LiDAR.
Prior work has proposed methods to create physical objects that evade LiDAR detection~\cite{tu2020physically,cao2019adversarial,zhu2021lidarattack} 
or evade sensor-fusion models~\cite{cao2021invisible,zhu24fusionattack,cheng2024fusion},
and a promising area of future work would be to extend our methodology 
to evaluate multi-modal or LiDAR-based attacks against driving agents in end-to-end settings.

\paragraph{Expanding threat models for adversarial ML.}
Prior work in adversarial ML commonly uses white-box (i.e. having access to the model weights) or black-box (i.e., only having access to model inputs and outputs) threat models~\cite{papernot2018sok}. 
In this work, we showed that, beyond using an ML model, agents use other modules that drastically affect attack success. 
We propose that threat models for adversarial ML should be expanded to include agent-specific modules, such as investigating how an attacker can leverage knowledge about 
a driving agent's code or configuration.
Additionally, future work could investigate methods for leaking an agent's code or configuration, 
perhaps using methods similar to those used for leaking ML model parameters~\cite{oliynyk2023stealing}.

\paragraph{Beyond CARLA-based robustness benchmarks for newly developed agents.}
In this work, we analyzed factors in driving agents that affect adversarial robustness
by leveraging standardized APIs for agent inputs (e.g., using the same camera) 
and outputs (e.g., controlling a vehicle with the same mechanics) from CARLA Leaderboard.
In a similar effort, the recently developed CARLA RAI Leaderboard compares driving agents based on 
bias and environmental impact~\cite{raileaderboard,omeiza2024rai}. 
Thus, we believe that CARLA serves as strong foundation for building a similar leaderboard for adversarial robustness.

However, we acknowledge that CARLA is not the only standardized platform available for autonomous driving;
platforms such as Baidu Apollo~\cite{fan2018baidu}, Autoware~\cite{kato2018autoware}, and PASS~\cite{shen2022pass}
each serve as high-fidelity platforms for testing and deploying models in autonomous vehicles.
However, due to their high-fidelity nature, these platforms pose a much greater barrier to entry compared to CARLA:
these platforms are designed for researchers to implement their own self-driving vehicles 
with all its relevant hardware and associated modules~\cite{jung2025open}. 
As a result, prior work that uses these platforms to evaluate the security of ML 
for autonomous driving must create their own customized system and scenarios 
for evaluation~\cite{cao2021invisible,zhu24fusionattack,wang2023does,Kim22,Huai23}.
These platforms can be used to demonstrate that a specific, well-crafted attack on a high-fidelity 
autonomous driving system is realistic and practical; 
in contrast, our work supports modularized, configurable evaluations of attacks on a variety of 
ML models used in driving agents, with a far lower barrier to entry.
A promising area of future work would be to extend environments like the CARLA Leaderboard
to support systematic evaluations of newly submitted models and agents in the context of 
higher fidelity systems based on Apollo, Autoware, or PASS.

%% file: conc.tex
\label{sec:conc}

In this work, we evaluate adversarial examples against driving agents submitted to the CARLA Leaderboard. 
We find that, beyond the ML models contained in the agents themselves, other modules used in driving agents can affect attack outcomes. 
For example, we show that an adversarial example can mislead an ML model to predict a left turn, but the attack is mitigated by the agent's GPS-based rules.
We propose directions for further investigations of agent-specific factors that affect robustness, 
and we recommend creating a standardized leaderboard for evaluating adversarial examples against driving agents, 
which can be used both to develop stronger attack strategies and build more robust driving agents. 

%% file: appendix.tex
\input{tables/app-experiment-params.tex}

\section{Additional experimental details}
\label{app:setup}

In this section, we provide additional details for how we define adversarial patches, and how we optimize patch values during patch generation.

In \tabref{table:parameters}, we list the parameters used to define patches for our attacks. 
For each agent and attack goal, we define the size of the patch (in pixels) and the size of the patch when streamed to an object in CARLA (in meters). 
We also support creating multiple patches that perform a single attack. We note that each agent uses a different number of cameras with a different input resolution, 
so evaluating attacks across agents with a standard patch size is unlikely to be effective. 
In general, we find that agents with higher input resolutions can be attacked with higher-resolution patches, 
and we find that stopping attacks can be performed with patches that are smaller than those required for steering attacks.
As a part of our work-in-progress, we are continuing to explore how our selected parameters affect attack success.

Next, we provide additional details for how we performed patch optimization in our experiments. 
When collecting images for patch optimization, we drive each agent through our target location in CARLA several times---five times for Road~\#5 and eight times for Turn~\#2. 
We observe a higher variance within driving scenarios that involve turns and additional data helps to capture this variance.
In our patch projection algorithm, we sample the linear scaling factor uniformly from $\mathbb{U}(0.9, 1.1)$ for color jitter, 
and we apply Gaussian blurring using a kernel size of 3 and $\sigma \sim \mathbb{U}(0.6, 1.0)$.
Based on the loss function defined in \secref{framework:attacks}, we weigh our loss terms with the following coefficients: $\alpha=2.0,\beta=3.0,\gamma=15.0,\eta=250.0,\delta=0.1,\epsilon=50.0$. 
To optimize our loss function, we use 500 epochs of AutoPGD, which dynamically tunes the step size during optimization. We restrict the step size from 4--128, optimizing on RGB values from 0--255.

For more sophisticated attacks, such as steering attacks, we modify the dataloader to dynamically adjust the input sequence, 
progressively revealing more of the route based on the patches' performance on the observed section. 
This iterative approach ensures that the attack remains effective during the critical initial frames where the trajectory should start deviating from the benign case. 
By prioritizing robustness in these early frames, we maximize the attack's ability to influence long-term vehicle behavior. 
Additionally, this method allows us to assess how the attack's effectiveness evolves over time, enabling a more systematic evaluation of our attack configurations.

%% file: tables/app-experiment-params.tex
\begin{table}[b!]
	\centering
	\caption{We list the parameters used to define patches for each attack in our experiments.}
	\label{table:parameters}
	\begin{tabular}{r|rrrr}
	\toprule
        \textbf{Agent} & \makecell[r]{\textbf{Attack}\\\textbf{Goal}} & 
            \makecell[r]{\textbf{Defined}\\\textbf{Size}} & \makecell[r]{\textbf{Streamed}\\\textbf{Size}} & \textbf{\#Patch} \\
    \midrule
        \multirow{2}{*}{TCP} & Stopping & 16px$^2$ & $4m^2$ & 1 \\
        & Steering & 18px$^2$ & $6m^2$ & 2 \\
    \midrule    
        \multirow{2}{*}{NEAT} & Stopping & 12px$^2$ & $4m^2$ & 1 \\
        & Steering & 12px$^2$ & $6m^2$ & 2 \\
    \midrule        
        Rails & Stopping & 12px$^2$ & $4m^2$ & 1 \\
    \bottomrule
	\end{tabular}
\end{table}